\documentclass[a4paper]{article}
\usepackage[utf8]{inputenc}
\usepackage{authblk}
\usepackage{setspace}
\usepackage[margin=1.25in]{geometry}
\usepackage{graphicx}
\graphicspath{ {./figures/} }
\usepackage{subcaption}
\usepackage{amsmath}
\usepackage{xcolor}
\usepackage[english]{babel}
\usepackage[utf8]{inputenc}
\usepackage{multicol}

\usepackage[style=numeric,sorting=none]{biblatex}
\title{Intrinsic time resolution of a hexagonal BC404 scintillator: A Monte Carlo study of interaction geometry and optical reflectivity}

\author[1,2*]{C. H.~Zepeda-Fern\'andez}
\author[2]{L. F. Rebolledo-Herrera}
\author[2]{E. Moreno Barbosa}

\affil[1]{Secretaría de Ciencia, Humanidades, Tecnología e Innovaci\'on}
\affil[2]{Facultad~de~Ciencias~F\'isico Matemáticas,~Benem\'erita~Universidad~Aut\'onoma~de~Puebla}

\affil[*]{Address correspondence to: hzepeda@fcfm.buap.mx}

\date{}

\begin{document}

\maketitle

\begin{abstract}

Time resolution is an important parameter in particle detectors, as it determines their ability to distinguish between two temporally separated events. It can be decomposed into electronic and intrinsic contributions, the latter being influenced by scintillation, optical transport, detector geometry, particle type and energy, and the sensitive area of the photosensor. In this work, a Monte Carlo simulation based on Geant4 is used to investigate the intrinsic time resolution of a hexagonal BC404 plastic scintillator. The analysis includes the evaluation of the optical-photon distribution over the hexagonal surfaces and the characterization of the intrinsic time resolution for different interaction configurations. The intrinsic time resolution is investigated using muons as cosmic-ray proxies with energies ranging from 4~GeV to 1~TeV. In addition, the effect of surface reflectivity on optical-photon transport and timing performance is studied for reflectivities ranging from 0\% to 100\%. The results show that the intrinsic time resolution exhibits little dependence on the incident muon energy, while a clear dependence on the interaction position is observed. Furthermore, although the number of detected photons varies with the interaction position and surface reflectivity, the intrinsic time resolution is not solely determined by photon statistics, indicating that optical transport and geometrical effects play an important role in the timing performance. Across the investigated configurations, intrinsic time resolutions ranging from approximately 15 to 100~ps are obtained. These findings highlight the importance of the interaction position, optical transport, photon statistics, and surface reflectivity in determining the timing performance of hexagonal plastic scintillators and provide guidance for the optimization of their optical properties and geometry for timing-sensitive applications.

\end{abstract}

{\bf keyworrds:} Plastic scintillator; Intrinsic time resolution; interaction geometry; optical reflectivity; Geant4.

\section{Introduction}\label{introduction}

Time resolution (TR) is an important parameter in particle detectors, particularly in beam monitoring applications~\cite{BeBe, LEONARD2014235, SLUPECKI2022167021}. This parameter enables the discrimination of temporally separated events, allowing for a reconstruction of particle trajectories~\cite{TheCMScollaboration_2013}. Detectors with high temporal accuracy play an important role in time-of-flight measurements~\cite{STRAZZI2025170521, RANTANEN2024169710}, cosmic-ray detection, and high-energy physics experiments.  In high-energy physics, beam monitors are commonly used to identify and discriminate minimum bias events in beam--beam interactions, as well as background contributions such as beam--gas interactions. In heavy-ion collisions, these detectors are also employed for the reconstruction of physical observables related to event centrality. Typically, such systems are based on materials with high radiation tolerance, as scintillating crystals~\cite{ATLAS, LHCb,IPPOLITOV2002121, ROGAN2010267}. As an additional remark, beam monitoring applications are relevant across the medical, research, and industrial sectors. Beam monitors are typically based on solid-state detectors, such as cadmium telluride (CdTe), silicon (Si), and cadmium zinc telluride (CZT), or on scintillation materials, including lutetium yttrium oxyorthosilicate (LYSO) and sodium iodide (NaI). The choice of material directly influences the detector parameter, particularly the temporal resolution, which is essential for accurate beam characterization and timing measurements~\cite{Habib2015, Hsieh2016, Crespo2016, Iniewski2011}. However, alternative approaches based on plastic scintillators have been proposed for beam monitoring applications~\cite{ALLEN2003549, ALICECollaboration_2008, BeBe}.\\
Plastic scintillators are widely used in fast timing detectors due to their high light yield, short decay time, and relatively low cost compared to crystal-based detectors~\cite{Sharma2023_JPETEfficiency, Siwal_2025, ZHANG2024165247, vaneijk2002, PDG2020, LI2005449, yanagida2018}. When coupled with silicon photomultipliers (SiPMs), they provide compact and efficient detection systems~\cite{particles8040094, inproceedings, STEINBERGER2019185, 9875794}. The use of the fast output of a SiPM provides a significantly improved temporal response due to its intrinsically short rise time, making it particularly suitable for applications requiring precise timing measurements. Although this output represents only a fraction of the total collected charge compared to the standard (slow) output, it has been demonstrated that the deposited charge can be reliably reconstructed by integrating the fast signal. Then, it is possible to preserve the charge information while achieving superior timing performance, effectively combining fast response and accurate charge estimation within a single readout scheme, as reported in the literature~\cite{Zepeda-Fernández_2020}.\\
In these systems, the TR depends on several factors, including the scintillation process, optical photon transport, detector geometry, photosensor characteristics, and readout electronics. This motivates the definition of the intrinsic time resolution (ITR), which excludes contributions from the electronic readout and data acquisition system. For example, for a telescope-type experimental setup composed of two detectors, {\it A} and {\it B}, where the ITR of detector {\it B} is known ($\sigma_{B}$) and that of detector {\it A} ($\sigma_{A}$) is unknown, the TR of the system ($\sigma_{\mathrm{TR}}$) can be expressed as:
\begin{equation}
\sigma_{\mathrm{TR}}^2 = \sigma_{A}^2 + \sigma_{B}^2 + \sigma_{\mathrm{elec}}^2,
\end{equation}
where $\sigma_{\mathrm{elec}}$ represents the contribution from the electronics. This method has been previously used to determine the ITR, both experimentally and through Monte Carlo simulations, of hexagonal plastic scintillator detectors~\cite{BeBe}. The transport of scintillation photons within the detector, which depends on the interaction geometry and the optical properties of the scintillator surfaces, can significantly influence the timing response. In particular, the reflectivity of the surrounding material affects the probability that optical photons reach the photosensor and their corresponding arrival-time distribution. Assuming a constant reflectivity $R$ at each interaction with the surrounding surface, the probability that an optical photon survives after $N$ reflections can be expressed as $P=R^N$~\cite{Roncali2017}. Therefore, even small changes in surface reflectivity can substantially modify the number and timing of detected optical photons, particularly in detectors where photons undergo multiple reflections before reaching the photosensor.
This relationship shows that the survival probability decreases rapidly as the number of reflections increases, particularly for low reflectivity values. Consequently, the optical boundary conditions can significantly affect both the number of photons reaching the photosensor and their propagation paths within the scintillator. Since photons undergoing different numbers of reflections may travel different path lengths before reaching the photosensor, changes in reflectivity can also modify the temporal distribution of detected photons and, therefore, the ITR of the detector. Understanding this interplay between surface reflectivity, photon statistics, and optical transport is therefore essential for characterizing the timing performance of scintillation detectors.\\
A relevant example of a beam monitoring detector based on hexagonal segmentation is the proposed Beam--Beam monitoring detector (BeBe)~\cite{BeBe}, designed for high-energy heavy-ion collision in the  Multipurpose Detector (MPD)~\cite{MPDCollaboration2022} at the Nuclotron-based Ion Collider fAcility (NICA) complex of the Joint Institute for Nuclear Research (JINR). The BeBe detector is composed of an array of hexagonal BC404 plastic scintillator cells, in order to function as a trigger, then it needs a fast response time. Its geometry allows for efficient spatial coverage and provides sensitivity to key observables such as event centrality and minimum bias discrimination, making it a suitable design for high-granularity timing detectors. Therefore, understanding the ITR and the role of optical and geometrical effects in such geometries is essential for the optimization and performance evaluation of detectors like BeBe, making the study of ITR an important parameter of its design.\\
In this work, a Monte Carlo study based on the Geant4 toolkit is presented to investigate the ITR of a hexagonal BC404 plastic scintillator and its dependence on the physical and geometrical parameters governing the detector response. In general, the ITR can be regarded as a function of several parameters,
\begin{equation}
\sigma_{\mathrm{ITR}} =
f(E,\mathbf{r}_{\mathrm{int}},R),
\end{equation}
where $E$ is the energy of the incident particle, ${\bf r}_{int}$ is its interaction position, $R$ is the reflectivity of the surrounding surface, for a fixed scintillator geometry and material.\\
The methodology described in Section~\ref{methodology} is organized into two consecutive stages, in which the results obtained at each stage provide the basis for the subsequent analysis. In  {\textit Stage I}, the spatial distribution of optical photons within the scintillator and over its surfaces is characterized. The dependence of the ITR on the incident-muon energy and interaction position is investigated. Muons with different energies are considered at four interaction configurations, allowing the effect of the interaction geometry on the timing response to be evaluated. The resulting photon statistics are also analyzed to determine whether the number of detected photons alone can account for the observed timing performance.  Finally, in {\textit Stage II}, the optical boundary conditions are systematically varied by changing the surface reflectivity. This stage builds upon the previous results by investigating how modifications of the optical-photon transport affect both the number of detected photons and the ITR. The corresponding results are presented in Section~\ref{results} following the same sequence. First, the spatial distribution of optical photons is used to characterize the detector response and the relevant sensitive surface, 
followed by the study of its dependence on muon energy and interaction position. The relationship between photon statistics and timing performance is subsequently examined, and the effect of surface reflectivity on the optical transport and ITR is finally investigated. Section~\ref{discusion} presents a discussion of the results, focusing on the physical interpretation of the findings presented in Section~\ref{results}. Finally, the conclusions of the study are presented in Section~\ref{Conclusions}.\\
This analysis provides insight into the interplay between optical transport, photon statistics, and detector geometry in determining the ITR of plastic scintillators. These simulation results are particularly relevant for the optimization of fast-timing detectors with non-uniform optical transport and complex geometries, and provide a basis for future experimental validation.

\section{Methodology}\label{methodology}
The simulations in this study were carried out using the Geant4 toolkit~\cite{Geant4}, version {\it geant4-v11.1.3}, which provides a comprehensive framework for modeling particle--matter interactions. The relevant physics processes for the production and transport of optical photons were included, these comprise energy deposition by charged particles through ionization, scintillation light production and Cherenkov radiation. The subsequent transport of optical photons is modeled taking into account reflection, refraction, absorption and boundary interactions at material interfaces~\cite{Korpachev_2017}, optical photon transport is dominated by interactions within the scintillator and at its boundaries. The world volume was defined using \textit{G4\_AIR} from the Geant4 material database.  The detector geometry, optical properties, boundary conditions, and particle source configuration are described in the following subsections. 

\subsection{Detector Geometry}\label{sec:geometry}
\begin{figure}
    \centering
    \includegraphics[width=0.7\linewidth]{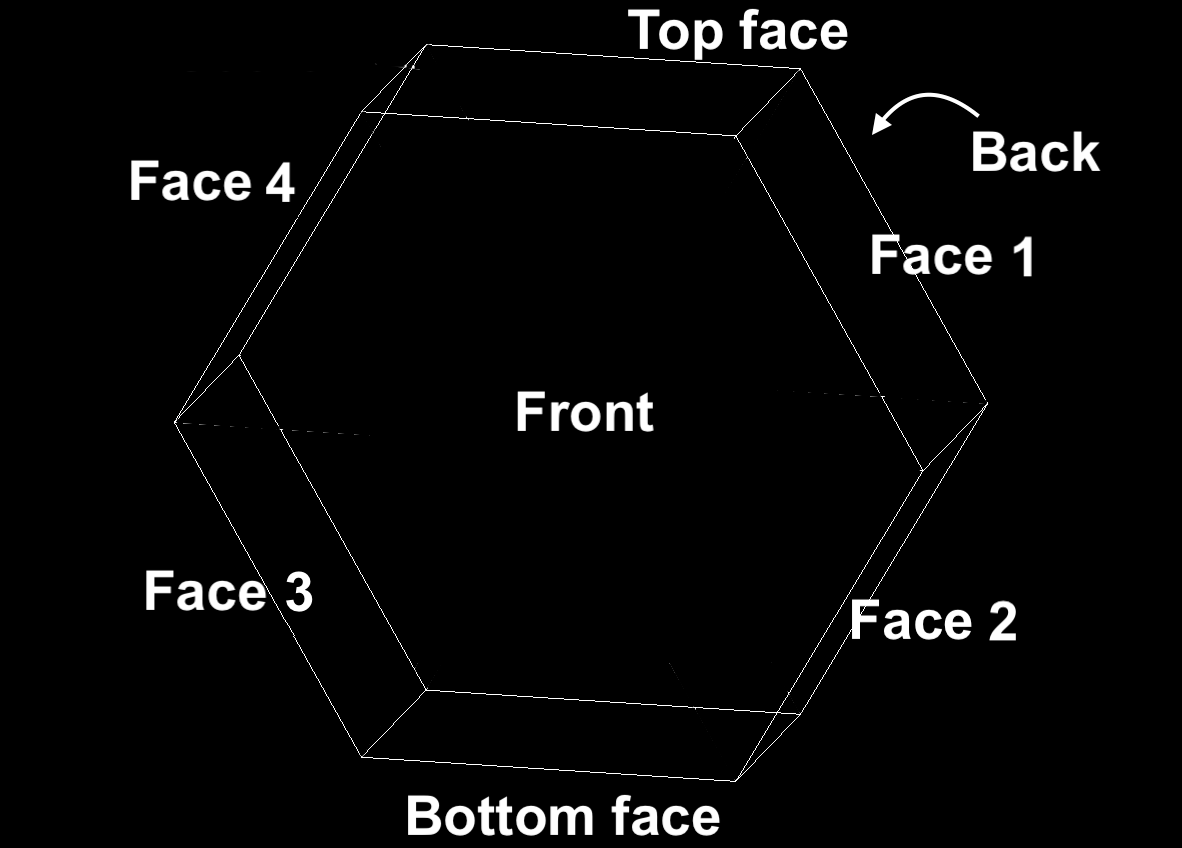}
    \caption{Schematic view of the hexagonal BC404 scintillator geometry showing the labeling of the eight faces and the position of the $6 \times 6$~mm$^2$ scorer on Top face.}
    \label{fig:geometry}
\end{figure}

A hexagonal BC404 scintillator was simulated with an apothem of 5~cm and a thickness of 2~cm. The geometry consists of eight faces: six lateral faces and two end faces (front and back). Figure~\ref{fig:geometry} shows the labeling convention adopted for each face. The sensitive area of a SiPM was emulated as a $6 \times 6$~mm$^2$ square detector (Scorer), modeled as a scoring surface. The resulting Scorer location is presented in Subsection~\ref{spatialdistribution}, based on the spatial distribution of the detected optical photons.  The scorer volume was defined using the \textit{G4\_GLASS\_PLATE} material from the Geant4 material database.
\subsection{Optical Properties and Boundary Conditions}\label{sec:optical}
\begin{figure}
    \centering
    \includegraphics[width=0.7\linewidth]{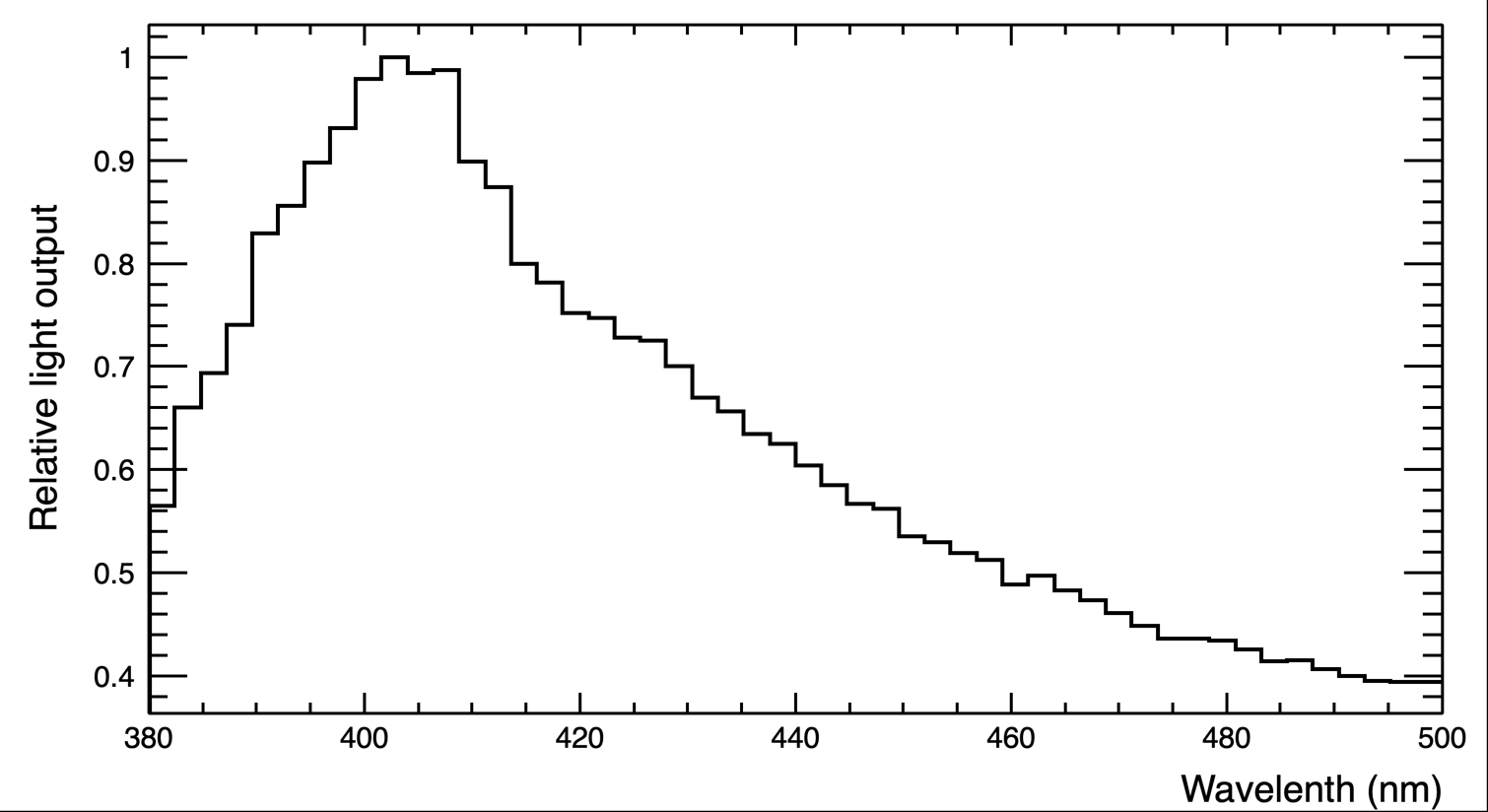}
    \caption{Emission spectrum of the BC404 scintillator used in the simulation.}
    \label{fig:spectrumemission}
\end{figure}

The base material of the BC404 plastic scintillator was defined using the {\it G4\_PLASTIC\_SC\_VINYLTOLUENE} material from the Geant4 material database. The following optical properties were assigned~\cite{BC404}:
\begin{itemize}
\item Light yield: 10,880 photons/MeV~\cite{lightouput}.
\item Refractive index: 1.58.
\item Attenuation length: 1.4~m.
\end{itemize}
The BC404 emission spectrum was implemented according to these parameters  and it shown in Figure~\ref{fig:spectrumemission}.\\
Two types of optical boundary conditions were defined:
\begin{itemize}
\item \textbf{Scintillator--world interface (dielectric--metal):} This surface was modeled with a reflectivity of 93\%, emulating a reflective wrapping such as Mylar.
\item \textbf{Scintillator--scorer interface (dielectric--metal):} To prevent double counting of photons, this surface was defined as 100\% absorbing. Consequently, the thickness of the scorer is irrelevant, as no photons are transmitted through it.\\
The objective of this study is to characterize the distribution of optical transport of the ITR, not to model the complete response of the SiPM.

\end{itemize}

\subsection{Analysis Strategy}\label{AnalysisStrategy}

The analysis was carried out in two stages, as described below.

{\bf Stage I Dependence on interaction position and particle energy}
 
 A total of 1,000 muon interactions with an energy of 1~GeV were simulated, with their interaction positions uniformly distributed over one of the hexagonal faces of the detector to avoid preferential sampling of specific photon-propagation paths. The spatial distribution of optical photons was investigated to characterize their transport within the hexagonal scintillator and to determine the photon yield over the different detector surfaces. For each event, the number of optical photons reaching each of the six lateral surfaces, as well as the front and back surfaces, was recorded. This analysis was used to determine the surface receiving the largest photon contribution and, consequently, to establish the Scorer location for the subsequent timing studies.
The dependence of the ITR on the incident-muon energy and interaction position was then investigated, using muons as proxies for cosmic-ray particles. Four interaction configurations were considered to evaluate the effect of the interaction geometry on the timing response: Top, Center, Bottom, and Random. In the Top and Bottom configurations, the muon trajectory was positioned close to and far from a selected lateral surface, respectively, with a distance of 0.5~mm from the corresponding surface. In the Center configuration, the trajectory passed through the central region of the scintillator, whereas in the Random configuration, the interaction position was randomly sampled. All configurations were simulated with a perpendicular muon trajectory. The incident-muon energies considered were 4~GeV, 6~GeV, 8~GeV, 10~GeV, 100~GeV, and 1~TeV. For each combination of incident energy and interaction configuration, 1,000 muon events were simulated.
For each event, the number of detected optical photons and the corresponding photon arrival time were recorded. These photon statistics were compared with the resulting ITR values to investigate whether the number of detected photons alone can account for the observed timing performance. The results of the energy and interaction position dependence are presented in Subsection~\ref{EnergyDependenceoftheITR}, while the relationship between photon statistics and timing performance is discussed in Subsection~\ref{PhotonStatisticsandTimingPerformance}.

{\bf Stage II Dependence on the optical boundary conditions}

The effect of the optical boundary conditions on the timing performance was investigated by varying the reflectivity of the surrounding surface. The reflector was modeled as an optical surface with reflectivity values ranging from 0\% to 100\%. Photons that were not reflected at the boundary were absorbed. For each reflectivity value, the number of detected optical photons and the corresponding arrival-time distributions were recorded. 
The resulting timing performance was compared across the different reflectivity conditions. The corresponding results are presented in Subsection~\ref{reflectorsurface}.

\section{Results}~\label{results}

\subsection{Spatial Distribution of Optical Photons}\label{spatialdistribution}

The spatial distribution of arrival optical photons to each hexagonal face was studied in order to evaluate the uniformity of light collection across the detector  and to identify an optimal location for the scorer. For this purpose, the number of photons reaching each face was recorded. The results show an overall uniform distribution of detected photons among the all faces, primarily driven by the geometrical configuration and boundary conditions of the system. However, the front and back faces exhibit a higher number of detected photons, as they are aligned with the muon direction, resulting in enhanced photon production along the particle trajectory. 
Table~\ref{tab:photcount} summarizes the average number of detected optical photons per event in each face of the hexagonal detector. This spatial behavior is expected to impact the ITR, as discussed in the following sections.

\begin{table}[htbp]
\centering
\caption{Average number of detected optical photons per event in each hexagonal face.}
\small
\begin{tabular}{lllllll}
\hline
\textbf{Hexagonal face} & \textbf{4 GeV} & \textbf{6 GeV} & \textbf{8 GeV} & \textbf{10 GeV} & \textbf{100 GeV} & \textbf{1 TeV} \\
\hline
Top face       & $260\pm$6 & 264$\pm$6 & 258$\pm$5 & 256$\pm$5  & 259$\pm$6 & 258$\pm$6\\
Face 1 & 263$\pm$6 & 263$\pm$6 & 262$\pm$6 & 262$\pm$6 & 260$\pm$6 & 263$\pm$6\\
Face 2 & 260$\pm$6 & 265$\pm$6 & 264$\pm$6 & 263$\pm$6 & 260$\pm$5 & 267$\pm$6\\
Bottom face & 267$\pm$6 & 268$\pm$6 & 266$\pm$6 & 269$\pm$6 & 264$\pm$6 & 267$\pm$6\\
Face 3 & 255$\pm$6 & 256$\pm$5 & 257$\pm$6 & 256$\pm$6 & 254$\pm$6 & 257$\pm$5\\ 
Face 4 & 251$\pm$5 & 252$\pm$5 & 250$\pm$5 & 249$\pm$5 & 250$\pm$5 & 253$\pm$5\\
Front    & 899$\pm$3 & 891$\pm$3 & 904$\pm$3 & 907$\pm$2 & 899$\pm$3 & 903$\pm$4\\
Back    & 895$\pm$3 & 892$\pm$3 & 898$\pm$2 & 903$\pm$2 & 907$\pm$2 & 911$\pm$11\\
\hline
\end{tabular}
\label{tab:photcount}
\end{table}

Furthermore, although the placement of the SiPM on the front or back faces would maximize photon collection, these regions are directly exposed to the incident radiation. In a realistic experimental setup, this could lead to sensor damage and the generation of spurious signals. For this reason, the study is conducted with the scorer positioned at the center of one of the lateral faces (Top face in Figure~\ref{fig:geometry}), providing a more robust and experimentally viable configuration. 
Based on these results of previous subsection, the four interaction points are renamed as:
\begin{itemize}
\item Near-scorer: Interaction close to the scorer.
\item Far-scorer: Interaction in the region furthest from the scorer.
\item Center-scorer: Interaction at the geometric center of the hexagon.
\item Random: Position randomly distributed over the surface of the scintillator. This approach allows the impact of spatial non-uniformities on the global ITR to be quantified.
\end{itemize}

Once the configuration has been defined, the ITR study is carry out.


\subsection{Dependence of the ITR on particle energy and interaction position}~\label{EnergyDependenceoftheITR}

\begin{figure}
    \centering
    \includegraphics[width=0.7\linewidth]{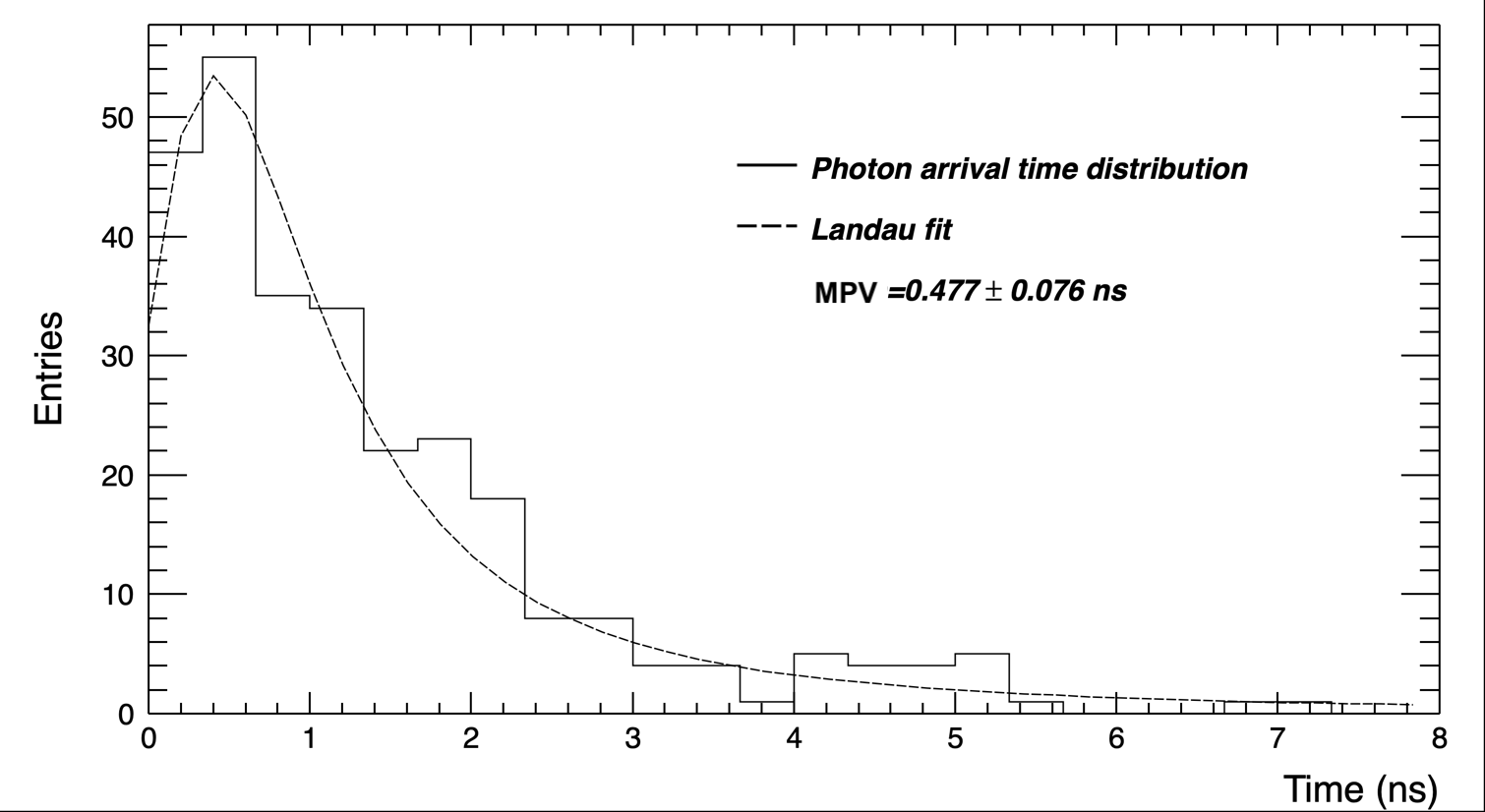}
\caption{Representative Landau fit to the optical-photon arrival-time distribution for a 10 GeV muon interacting at the center of the scintillator. The most probable value (MPV) obtained from the fit is used to characterize the timing response of each simulated event.}
    \label{fig:landaufit}
\end{figure}

\begin{figure}
    \centering
    \includegraphics[width=0.7\linewidth]{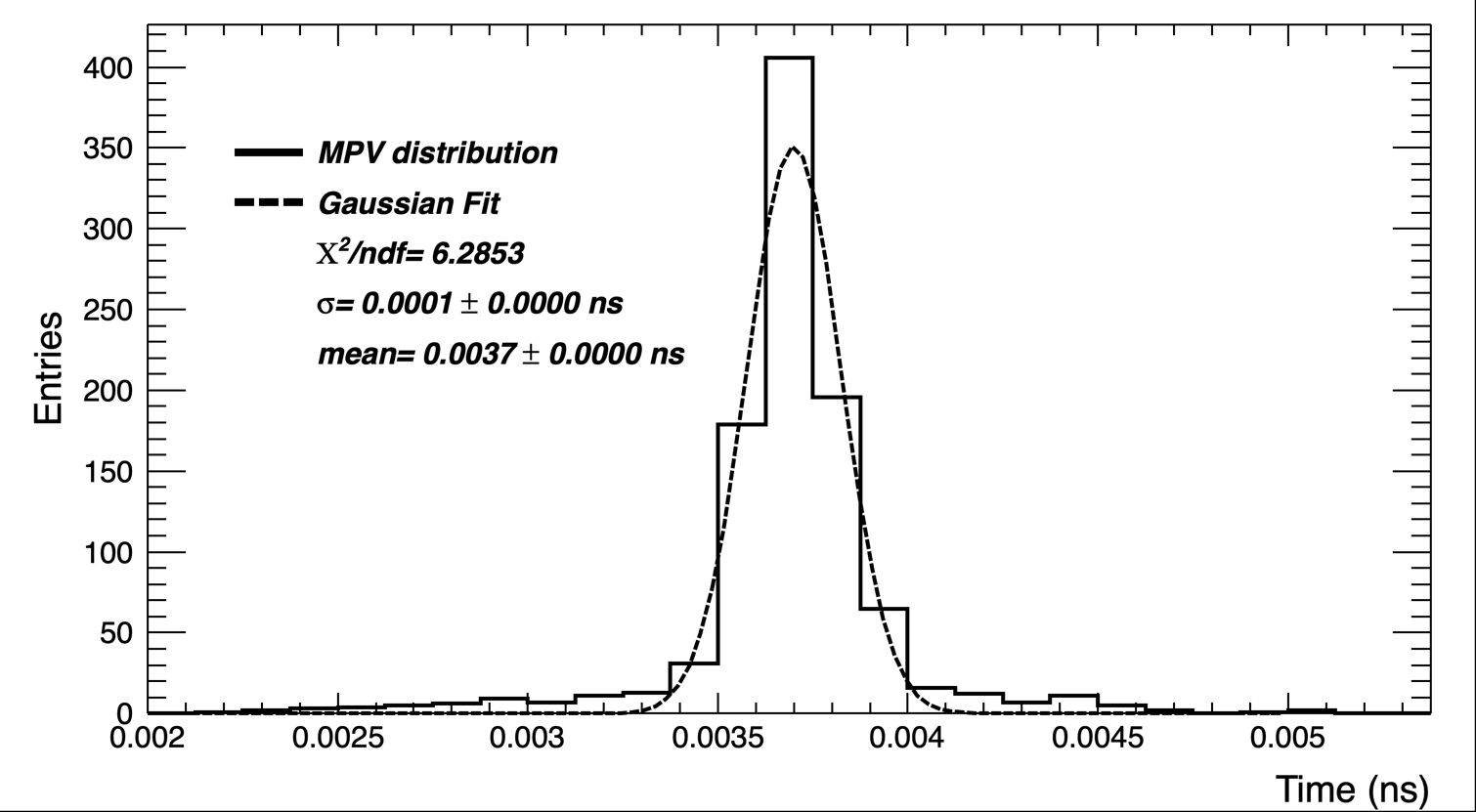}
    \includegraphics[width=0.7\linewidth]{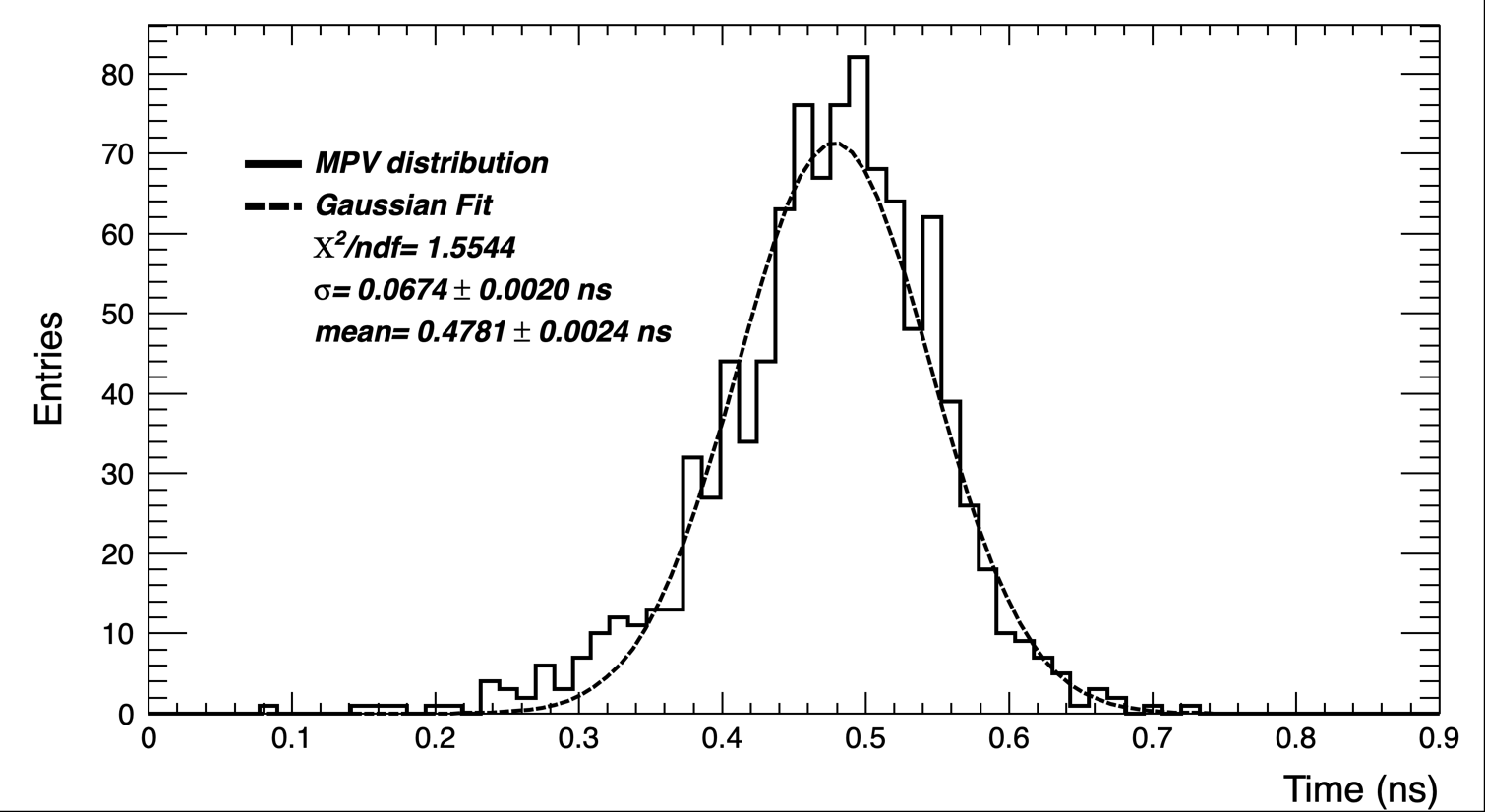}
    \includegraphics[width=0.7\linewidth]{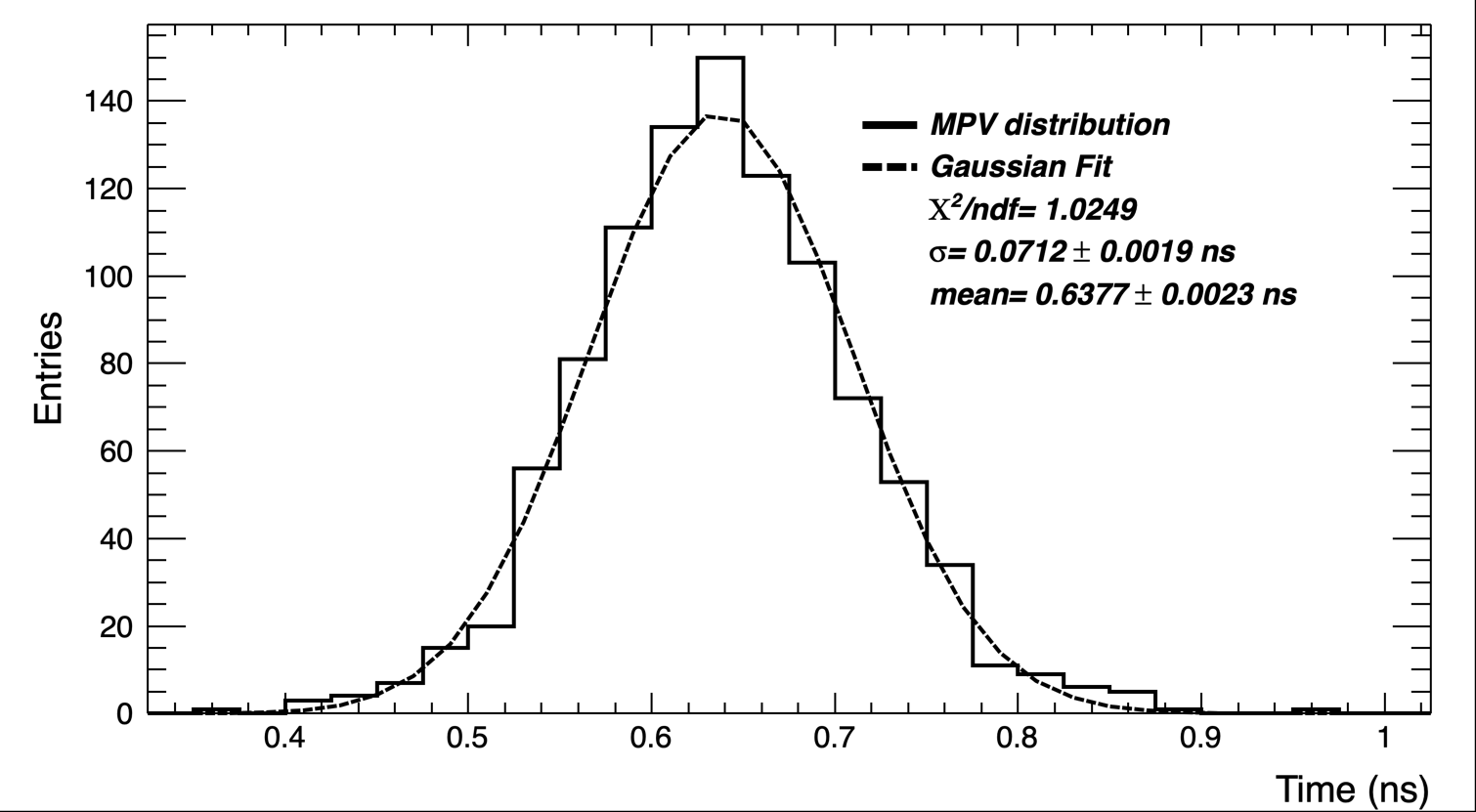}
    \includegraphics[width=0.7\linewidth]{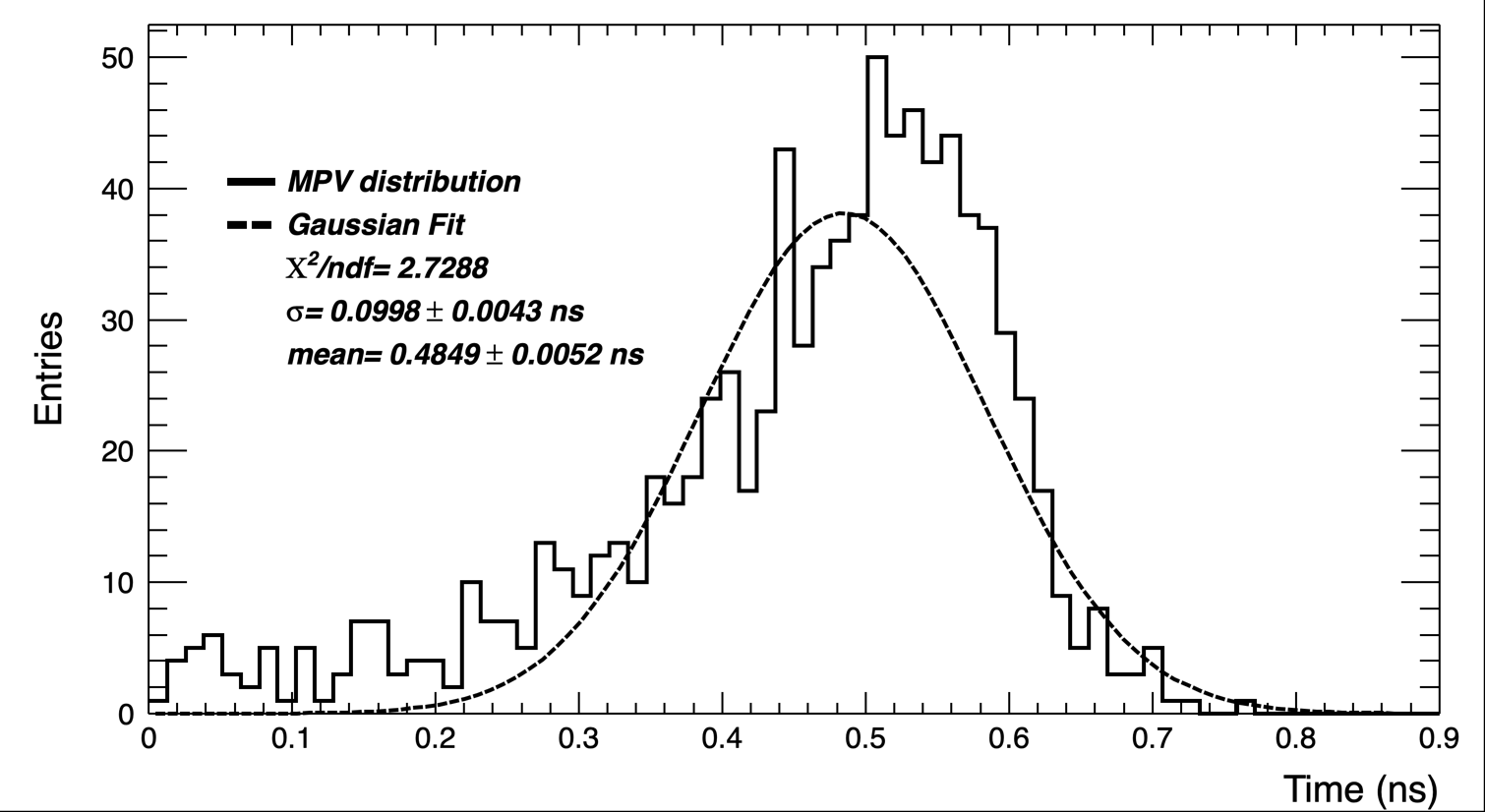}
\caption{Gaussian fits to the time distributions for different interaction positions at 10~GeV. The variation in width reflects the impact of photon transport and geometrical effects on the ITR.}
    \label{fig:gaussian_10GeV}
\end{figure}

As the scorer was located on Top face, for the near-scorer and far-scorer the interaction point was located 0.5~mm from the hexagonal edge surface.\\
The ITR was evaluated following the same procedure as in~\cite{BeBe,maque}, which consists of for each event  and configuration the arrival time distribution of optical photons to the scorer was obtained. Where an asymmetric shape is exhibited, due to photon propagation effects; therefore, the most probable value (MPV) was extracted from the respective Landau fit. The MPV of the photon arrival-time distribution was used as a representative estimator of the detector timing response. This choice assumes that the number of detected optical photons is sufficiently large to produce a measurable SiPM signal, such that the timing response is determined by the overall photon-arrival distribution rather than by a single detected photon. Using the earliest photon alone could result in a timing estimator dominated by an individual photon that may not be sufficient to produce a measurable detector signal under realistic experimental conditions. To illustrate this procedure, a fit distribution is shown in Figure~\ref{fig:landaufit}. The event-by-event distribution of the MPV was then constructed for each energy and interaction configuration. A Gaussian fit was considered and the standard deviation ($\sigma$) was taken as the ITR. Since $\sigma$ quantifies the statistical dispersion, it represents the minimum temporal uncertainty associated solely with the scintillator. As a representative example, Figure~\ref{fig:gaussian_10GeV} shows the Gaussian fits to the time distributions for the four interaction configurations at an incident muon energy of 10~GeV.

\begin{figure}
    \centering
    \includegraphics[width=0.7\linewidth]{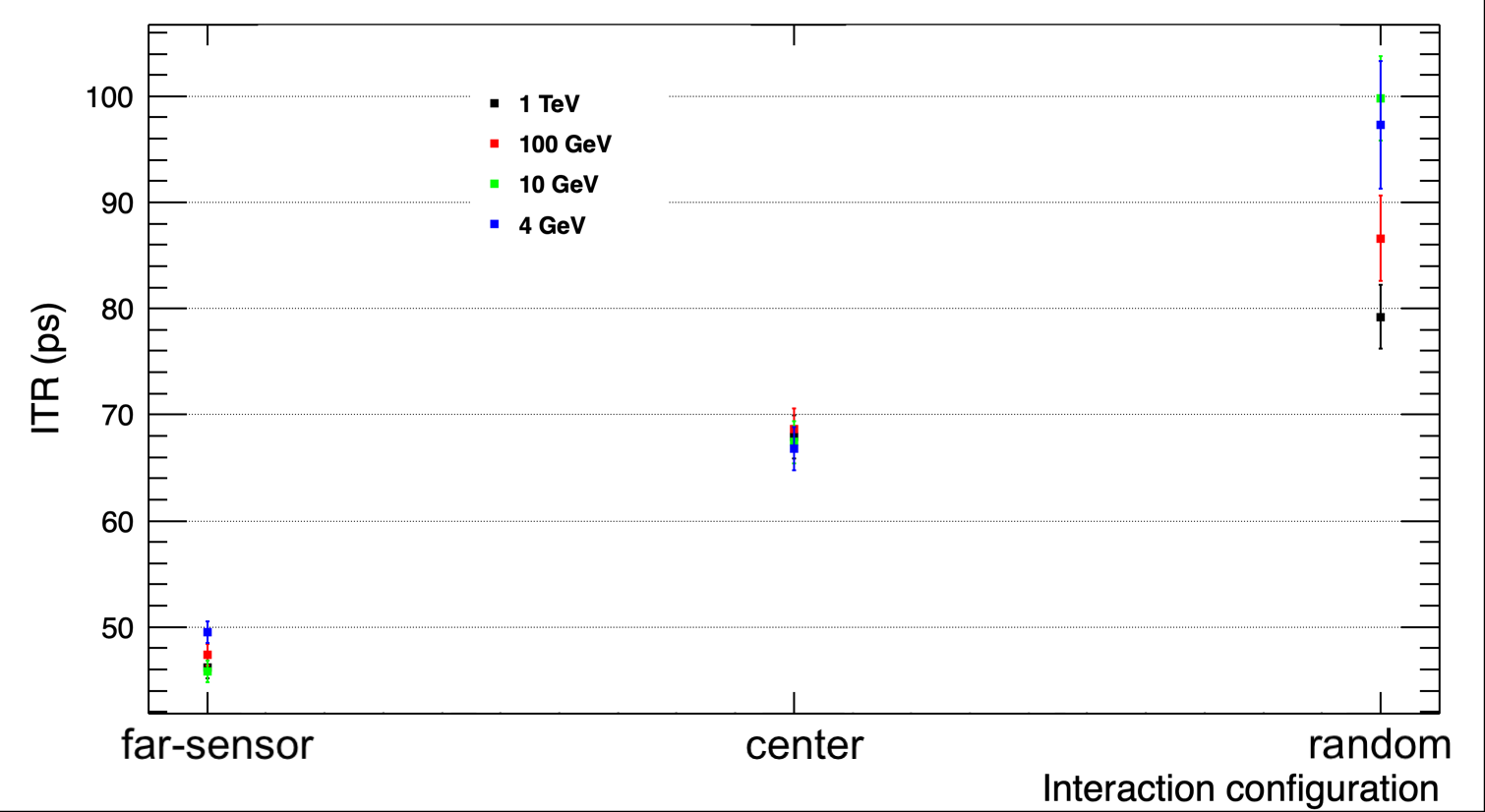}
\caption{ITR for different interaction configurations. The increased resolution observed for the random distribution highlights the dominant contribution of spatial non-uniformities compared to fixed interaction points.}
    \label{fig:intrinsictime}
\end{figure}

The near-scorer interaction exhibits a significantly narrower distribution, reflecting the minimal photon propagation path. In contrast, the center and far-scorer configurations show broader but distinct distributions, with the far-scorer configuration yielding the lowest ITR, followed by the center case, while the random interaction configuration exhibits the largest spread.  This behavior is consistently observed across all studied energies. Figure~\ref{fig:intrinsictime} shows the distributions of the ITR for all studied energies and for the center, far-scorer, and random interaction configurations. These results apparently confirm that the ITR is largely independent of the incident muon energy and is instead dominated by geometrical and optical transport effects. The near-scorer interaction exhibits an ITR of approximately 0.1~ps for all energies, due to the minimal photon propagation distance and the negligible contribution from optical transport and geometrical effects.  This trend highlights the role of photon transport: fixed interaction positions lead to more uniform optical paths, whereas a distributed interaction volume introduces a wider range of photon trajectories, increasing the temporal dispersion.\\
For the random interaction configuration, the arrival time distribution deviates from a Gaussian shape, exhibiting an asymmetric tail. In this case, the time resolution extracted from a Gaussian fit should be interpreted as an effective parameter, as it does not fully account for the contribution of non-Gaussian features in the distribution.\\
Having studied the relationship between energy and ITR, the results regarding the behavior of the ITR versus the number of detected photons are presented below.

\subsection{Relationship between photon statistics and timing performance}\label{PhotonStatisticsandTimingPerformance}

\begin{figure}
    \centering
    \includegraphics[width=0.7\linewidth]{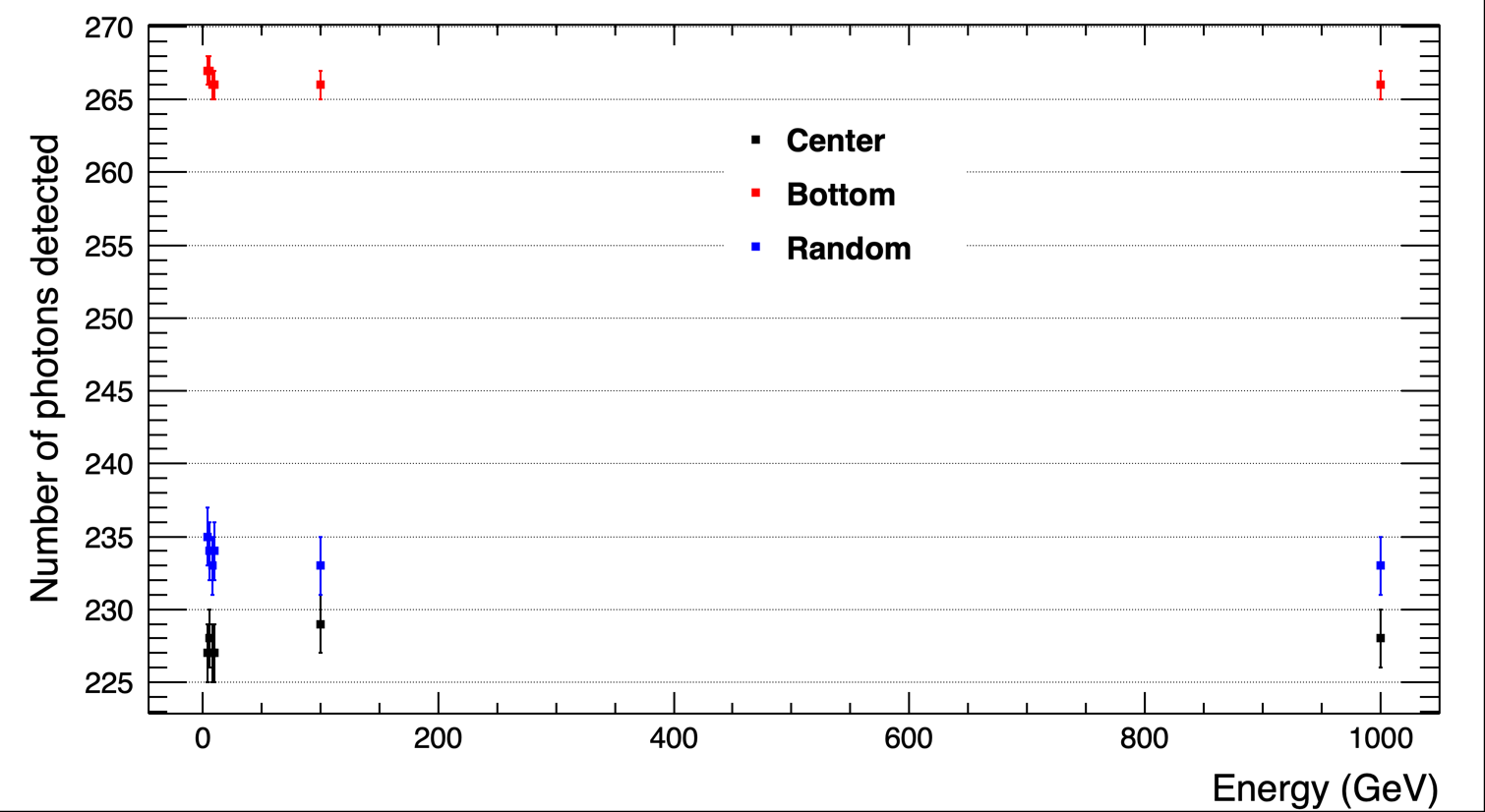}
\caption{Average number of detected optical photons as a function of the incident muon energy for different interaction configurations. The photon yield remains approximately constant across energies, with higher values observed for the far-scorer configuration due to geometrical effects.}\label{fig:photonsvsenergy}
\end{figure}

\begin{figure}
    \centering
    \includegraphics[width=0.7\linewidth]{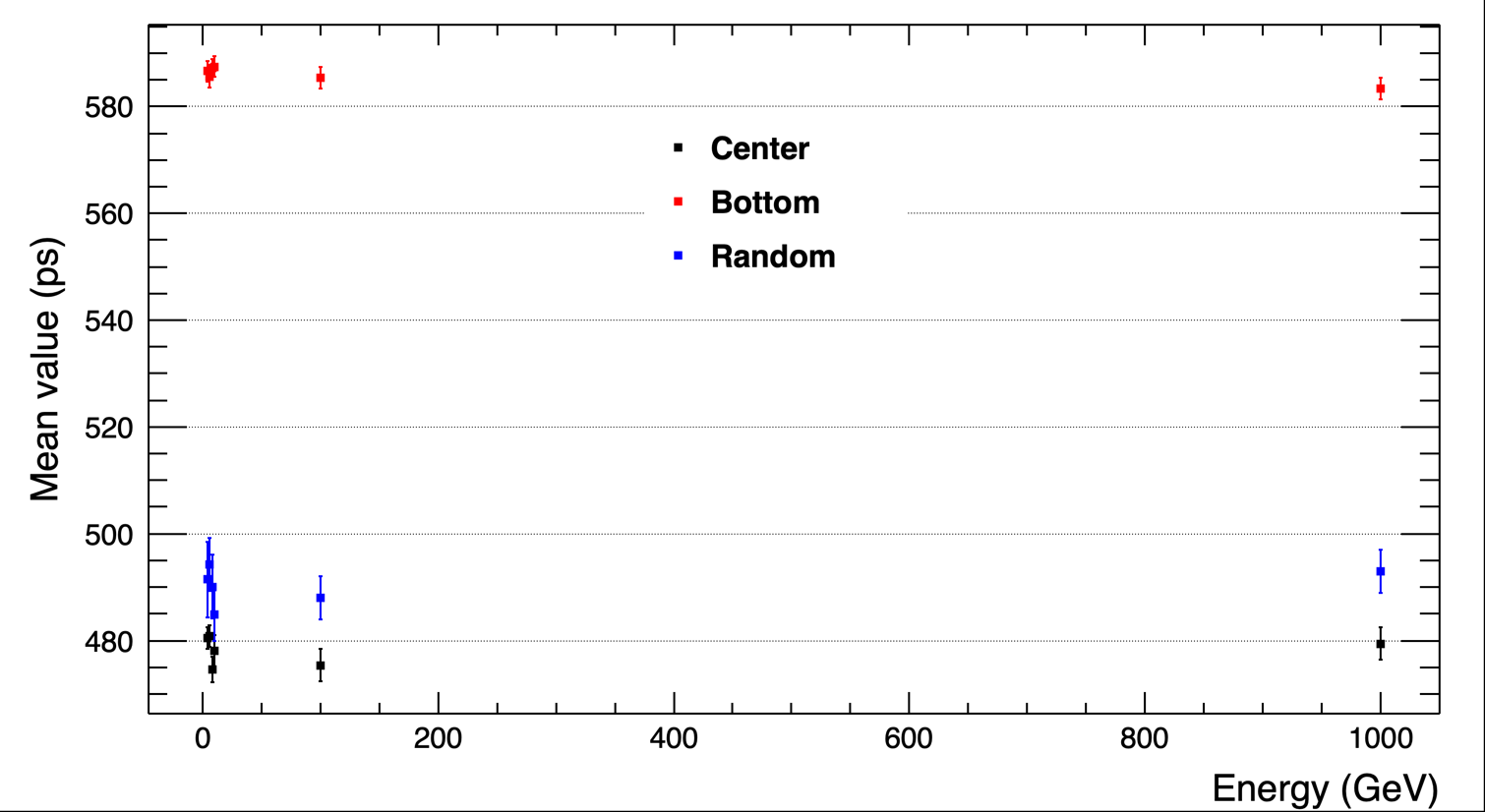}
\caption{Mean optical photon arrival time as a function of the incident muon energy for different interaction configurations. The mean remains approximately constant across energies, with higher values observed for interactions farther from the photosensor due to longer photon propagation paths.}\label{fig:meanvsenergy}
\end{figure}

The relationship between photon statistics and timing performance was investigated by analyzing the average number of detected optical photons for muons with energies ranging from 4~GeV to 1~TeV, considering the four interaction configurations previously defined. The results show a  dependence of the photon yield on the interaction position, as it is shown in Figure~\ref{fig:photonsvsenergy}. A larger number of detected photons is observed for the far-scorer  configuration, with an average of 266$\pm$1 photons per event, compared to about 229$\pm$2 and 235$\pm$2 photons for the center and random configurations, respectively.  In contrast, the near-scorer  configuration exhibits a significantly higher photon yield, with approximately $685 \pm 3$ detected photons per event across all energies. This enhancement is expected due to the minimal photon propagation distance and the reduced probability of absorption losses, as photons are generated in close proximity to the photosensor. However, despite this higher photon yield, the corresponding improvement in ITR is limited, reinforcing the conclusion that photon statistics alone do not determine the timing performance. This behavior can be attributed to the combined effect of photon production along the particle trajectory and multiple internal reflections within the scintillator. Due to the high reflectivity of the boundaries, photons generated at larger distances can still reach the photosensor, leading to an enhanced photon collection efficiency in the far-scorer configuration. Despite these differences in photon yield, the ITR is lowest for the far-scorer configuration and increases for the center case, demonstrating that timing performance is governed by optical transport and geometrical effects rather than photon statistics. These results demonstrate that increasing the number of detected photons does not necessarily lead to an improvement in ITR in this configuration, highlighting the dominant role of optical propagation and spatial effects in determining the detector timing performance. This behavior is further illustrated in Figure~\ref{fig:meanvsenergy}, which shows the mean value of the MPV distribution for different interaction configurations as function of the energy. A clear spatial dependence is observed: the mean value is lowest for the center configuration, increases for the random distribution, and reaches its highest value for the far-scorer  case. For the near-scorer  configuration, the mean arrival time is approximately 3.7~ps for all energies. This lower value is expected, as this configuration corresponds to interactions occurring in close proximity to the scorer, resulting in shorter photon propagation distances. This trend reflects the variation in photon propagation distances. Interactions occurring farther from the photosensor result in longer photon travel paths and therefore larger arrival times. The random configuration exhibits intermediate values, as it includes both near- and far-interaction events, while the center configuration remains closer on average to the scorer, leading to smaller mean arrival times. This result further emphasizes the role of spatial effects, showing that the configuration with the lowest ITR (far-scorer) does not correspond to the earliest photon arrival time, indicating that the timing spread and the mean arrival time are governed by different aspects of optical transport.

\begin{figure}
   \centering
    \includegraphics[width=0.7\linewidth]{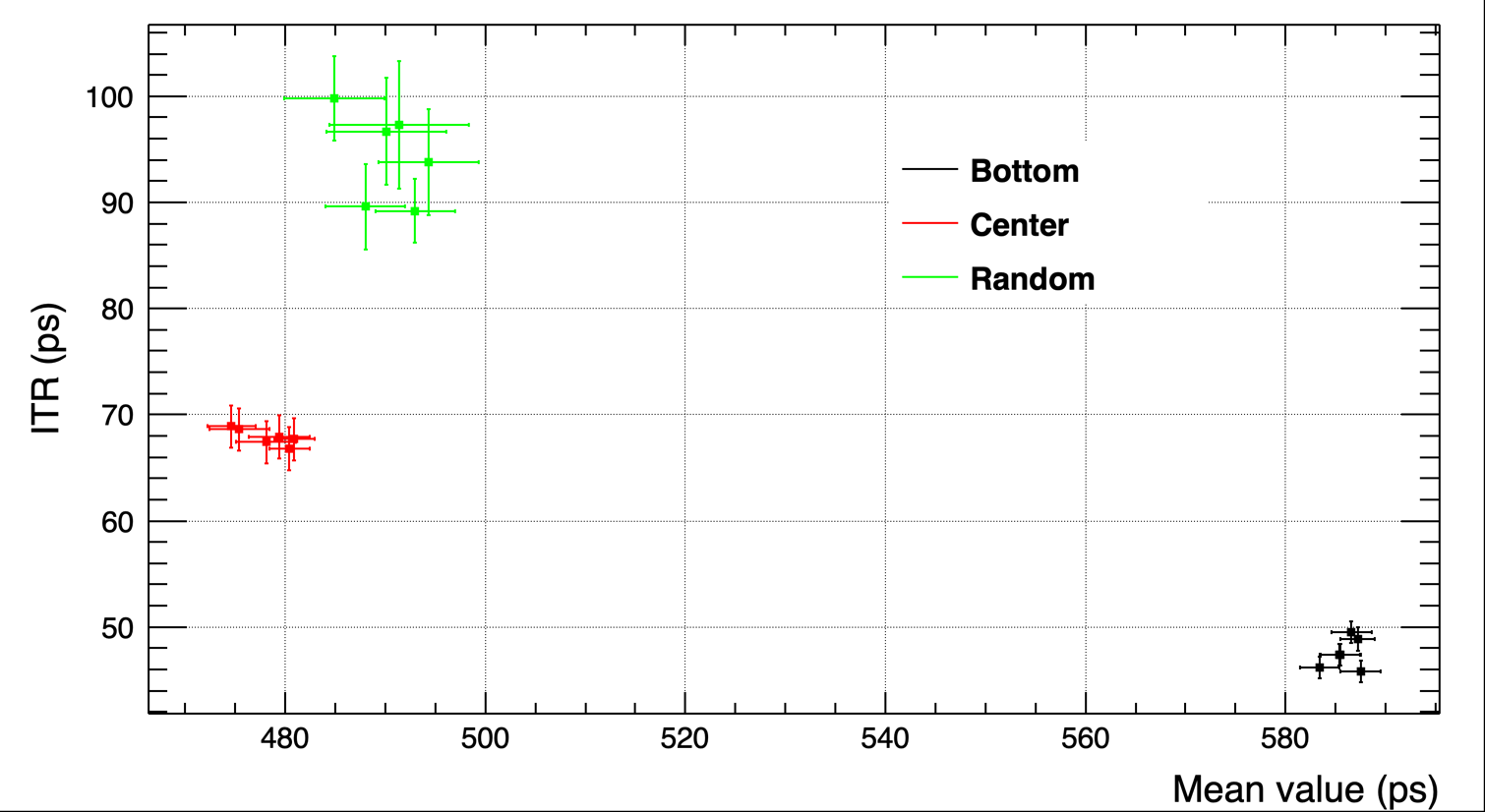}
\caption{ITR versus mean number of detected optical photons for the bottom, center, and random interaction configurations at different muon energies.}\label{fig:ITRVsPhot}
\end{figure}

Finally, to summarize the results discussed above, Fig.~\ref{fig:ITRVsPhot}  shows the ITR as a function of the mean number of detected optical photons for the different interaction configurations and particle energies. The bottom configuration exhibits the highest photon yield, with an average of more than 580 detected photons, and also provides the best timing performance, with an ITR below 50 ps. In contrast, the center and random configurations produce comparable numbers of detected photons, with mean values of approximately 480–500 photons, but exhibit markedly different ITR. The center configuration yields an ITR below 70 ps, whereas the random configuration results in an ITR of approximately 90–100 ps. Therefore, the number of detected photons alone does not fully determine the ITR. The difference between the center and random configurations, despite their similar photon statistics, indicates that the temporal distribution of the detected photons and, consequently, the distribution of their optical path lengths also play a significant role in determining the timing performance.

\subsection{Effect of optical reflectivity on the ITR}\label{reflectorsurface}
As described in the methodology, a reflectivity of 93\% was considered as the reference condition in this work. To investigate the dependence of the ITR on the optical reflectivity, additional simulations were performed using reflectivities of 80\%, 60\%, 40\%, and 20\%. A further case with 0\% reflectivity was also considered, in which the optical boundary is governed exclusively by the reflection and transmission processes at the interface between the two Geant4 media according to the Fresnel equations. For each reflectivity value, 1,000 muon events were simulated with an incident energy of 4~GeV and a center interaction position. The incident energy and interaction position were kept fixed throughout this study, so that the effect of the surface reflectivity on the ITR could be evaluated independently.\\
Under this condition, the detected photons are predominantly those reaching the scorer through relatively direct optical paths, while photons requiring multiple reflections at the surrounding surfaces have a lower probability of reaching the scorer. As the reflectivity decreases, the average number of detected optical photons also decreases. 
The average number of detected optical photons for each reflectivity condition is reported in Table~\ref{tab:averagephotons}.

\begin{table}[htbp]
\centering
\caption{Average number of detected optical photons for different surface reflectivity values.}
\small
\begin{tabular}{| c | c | c | c |}
\hline
Reflectivity percentage & number of \\
 percentage & Optical photons detected\\
\hline
0  & 280$\pm$2\\
20 & 9$\pm$1\\
40 & 20$\pm$1\\
60 & 42$\pm$1\\
80 & 105$\pm$1\\
93 & 274$\pm$2\\
100 & 1052$\pm$6\\
\hline
\end{tabular}
\label{tab:averagephotons}
\end{table}

\begin{figure}
   \centering
    \includegraphics[width=0.7\linewidth]{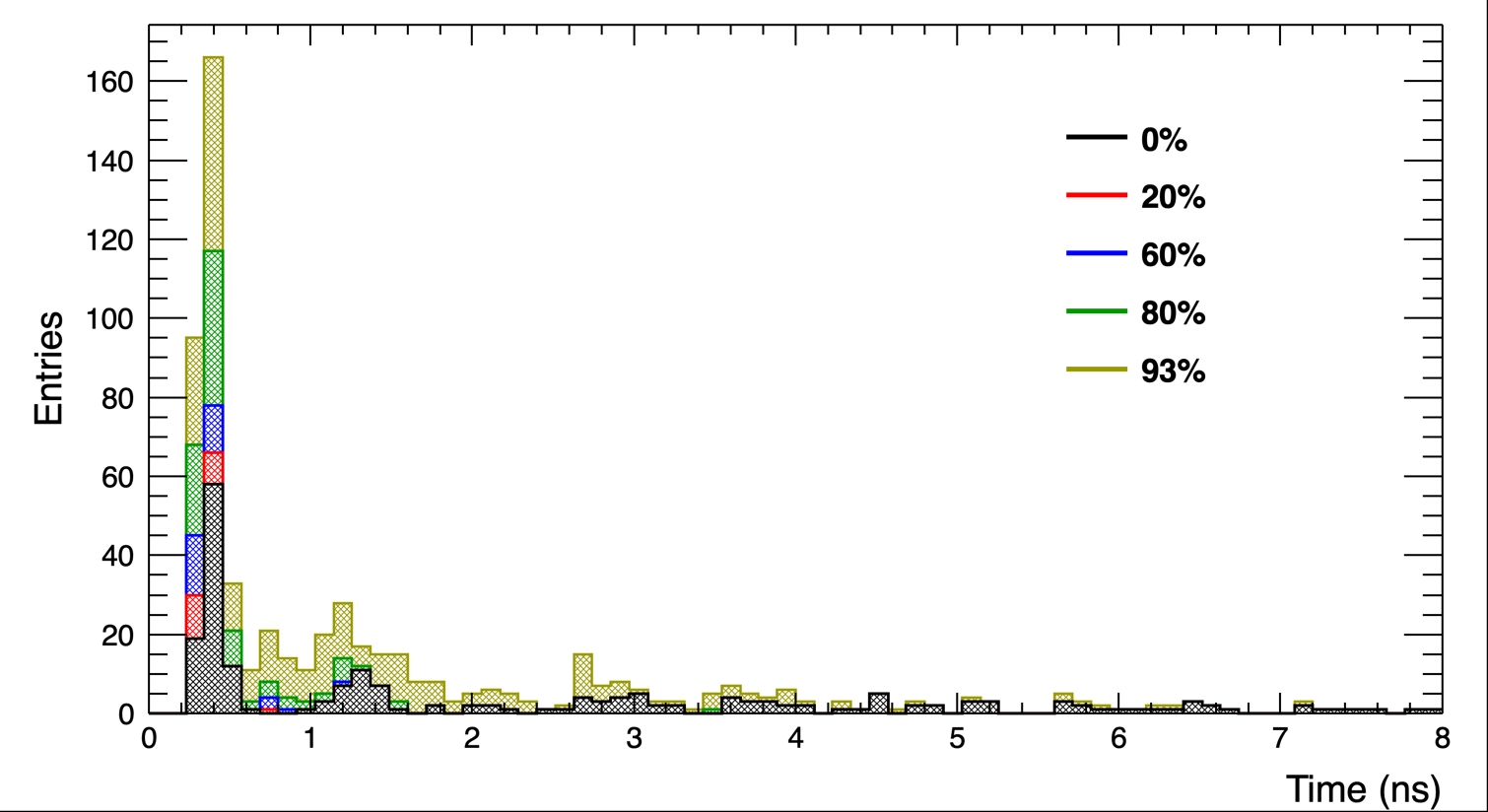}
\caption{Distribution of optical-photon arrival times for a single event for surface reflectivities of 0\%, 20\%, 60\%, 80\%, and 93\%. The distributions illustrate the effect of the optical boundary conditions on the temporal distribution of detected photons..}\label{fig:percentagedistribution}
\end{figure}

For the standard analysis, the MPV is obtained from the Landau fit to the optical-photon arrival-time distribution. This approach provides an uncertainty associated with the fitted MPV, allowing the statistical uncertainty of the extracted timing parameter to be quantified. Thus, each simulated event provides an MPV together with its corresponding fitting uncertainty. The same MPV-based procedure could also be applied to the 80\% and 0\% reflectivity configurations, yielding ITR values of 45.3$\pm$1.4~ps and 14.9$\pm$0.3~ps, respectively. The 0\% reflectivity configuration yields the lowest ITR, which can be attributed to the reduced contribution of photons undergoing multiple reflections at the scintillator boundaries. However, for the 60\%, 40\%, and 20\% configurations, the MPV-based procedure could not be directly applied because the arrival time distributions become progressively narrower and less compatible with a Landau shape, making the extraction of the MPV from the fit less reliable. Figure~\ref{fig:percentagedistribution} shows, as an example, the optical photon arrival time distributions obtained for a single event at each reflectivity value. Therefore, the MPV was obtained directly from the corresponding histogram. In this case, the extracted MPV depends on the chosen bin width, introducing a systematic uncertainty associated with the histogram-based extraction procedure. To evaluate this effect, two binning configurations were considered for the 1,000 simulated events, using 90 and 100 bins. These values were selected to reduce the coarseness of the histogram and to evaluate the dependence of the MPV determination on the binning choice.

For each bin width, the MPVs obtained from the 1,000 simulated events were used to construct the corresponding MPV distribution. Each distribution was fitted with a Gaussian function, from which the mean and standard deviation were obtained. The standard deviation $\sigma_1$ and $\sigma_2$, obtained for the two bin widths provides two independent estimates of the ITR, each with its corresponding statistical uncertainty from the Gaussian fit. Since neither bin width is preferred, the final ITR value was obtained as the average of the two estimates,

\begin{equation}
\sigma_{ITR}=\frac{\sigma_1+\sigma_2}{2}
\end{equation}
The difference between the two was used to quantify the systematic uncertainty associated with the choice of bin width,
\begin{equation}
Err_{\sigma_{Sys}}=\frac{\sigma_1-\sigma_2}{2}
\end{equation}
 Finally, as the statistical errors are independent, the total statistical error was calculated through the statical error corresponding to each bin ($Err_{\sigma_{i_{Sta}}}$, for i=1,2),
 \begin{equation}
Err_{\sigma_{Sta}}=\frac{\sqrt{Err_{\sigma_{1_{Sta}}}^2+Err_{\sigma_{2_{Sta}}}^2}}{2} .
 \end{equation}
The resulting ITR values are reported in Table~\ref{tab:ITRMPV}, where the first and second uncertainties represent the statistical and systematic contributions, respectively.

\begin{table}[htbp]
\centering
\caption{ITR as a function of the surface reflectivity. The first and second uncertainties correspond to the statistical uncertainty and the systematic uncertainty associated with the binning choice, respectively.}
\small
\begin{tabular}{| c | c | c | c |}
\hline
Reflectivity percentage & ITR (ps) \\
 percentage & \\
\hline
60  & 47.77$\pm$ 3.6$\pm$0.81\\
40 & 43.82$\pm$0.01$\pm$0\\
20 & 19.19$\pm$0.01$\pm$0\\
\hline
\end{tabular}
\label{tab:ITRMPV}
\end{table}

 \section{Discussion}\label{discusion}

\subsection{Photon statistics versus optical transport}
For each interaction configuration, a consistent ITR was obtained over the investigated muon-energy range from 4~GeV to 1~TeV, indicating a weak dependence of the timing resolution on the incident particle energy under the conditions considered. In contrast, the number of detected optical photons exhibited a dependence on the interaction configuration. The center and random configurations produced similar photon yields, both of which were lower than that obtained for the far-scorer configuration. Despite their comparable photon statistics, the center and random configurations exhibited different timing performances, satisfying
$\sigma_{\mathrm{ITR}}^{\mathrm{center}}<\sigma_{\mathrm{ITR}}^{\mathrm{random}}$.
This result demonstrates that the number of detected photons is not, by itself, a sufficient descriptor of the timing response. Instead, the temporal distribution of the detected photons also plays a significant role in determining the ITR, since photons following different optical paths can reach the photosensor at different times.
\subsection{Role of optical reflectivity}
According to the photon survival probability ($P=R^N$) the probability of photon survival depends on both the surface reflectivity (R) and the number of reflections (N). Increasing the reflectivity increases the probability that photons undergoing multiple reflections remain available for detection. However, these additional reflections also increase the diversity of optical path lengths and, consequently, the spread of photon arrival times. Therefore, an increase in photon yield does not necessarily result in an improved ITR. The results obtained in this work illustrate this behavior, showing that the timing performance depends non-monotonically on the surface reflectivity. In particular, the realistic reflectivity considered in this study provides a better ITR than the ideal 100\% reflectivity case. This result emphasizes that the optimization of the reflective properties of a scintillator should consider not only photon collection efficiency but also the temporal distribution of the detected optical photons.\\
This result illustrates that maximizing photon collection is not necessarily equivalent to minimizing the ITR. Although increasing reflectivity enhances photon survival, it also increases the contribution of photons undergoing multiple reflections and therefore broadens the distribution of optical path lengths and arrival times. From the perspective of optical transport alone, a low-reflectivity configuration may therefore provide a faster timing response. However, in a real experimental setup, reducing the reflectivity of the scintillator surfaces may be counterproductive, as it can increase the susceptibility of the detector to external light contamination and compromise the reliability of the optical readout. Therefore, the optimum surface treatment must consider a trade-off between timing performance, photon collection, and light shielding.
\subsection{Implications for detector design}
The results presented in this work provide useful guidelines for the design of fast-timing scintillation detectors, although they should ultimately be validated through experimental measurements. The simulations indicate that several parameters should be considered simultaneously when optimizing the detector timing response, including the photosensor position, scintillator geometry, surface reflectivity, uniformity of photon collection, and optical-photon yield. In particular, optimizing the number of detected photons alone may not be sufficient to achieve the best timing performance. Instead, the optical configuration should be designed to provide a favorable temporal distribution of photon arrival times while maintaining an adequate photon yield.

\section{Conclusions}\label{Conclusions}
A detailed Monte Carlo study of the ITR of a hexagonal BC404 plastic scintillator has been presented. The spatial distribution of optical photons was analyzed to characterize the uniformity of light collection over the detector surfaces. A nearly uniform photon distribution was observed among the lateral faces, whereas a higher photon yield was found at the front and back faces due to their alignment with the incident particle direction. Although these faces provide enhanced photon collection, their direct exposure to the incident radiation makes them less suitable for photosensor placement in realistic experimental configurations. Consequently, a lateral face was selected as a suitable and robust location for the photosensor.\\
The ITR was evaluated for muons with energies of 4~GeV, 6~GeV, 8~GeV, 10~GeV, 100~GeV, and 1~TeV, considering four interaction configurations: near-scorer, center, far-scorer, and a uniform random distribution. The energy deposition and average number of detected photons remained approximately constant over the studied energy range, resulting in only a weak dependence of the ITR on the incident muon energy. In contrast, a clear dependence on the interaction position was observed. The photon yield showed a moderate dependence on the interaction configuration, with the highest average number of detected photons obtained for the far-scorer configuration, followed by the random and center configurations. However, the timing performance did not directly correlate with the number of detected photons. In particular, the center and random configurations exhibited comparable photon statistics but significantly different ITR values. This demonstrates that photon statistics alone are insufficient to determine the timing performance.\\
The mean photon arrival time exhibited a clear dependence on the interaction position, increasing from the center to the random and far-scorer configurations, consistent with differences in the optical propagation distances. In contrast, the ITR followed a different behavior, with the lowest values obtained for the far-scorer configuration, followed by the center configuration, while the random configuration exhibited the largest temporal spread. These results demonstrate that the mean photon arrival time and its temporal spread are governed by different characteristics of optical-photon transport. Consequently, an earlier mean photon arrival time does not necessarily imply a better ITR.\\
The effect of the optical boundary conditions was further investigated by varying the surface reflectivity. The ITR exhibited a non-monotonic dependence on reflectivity. This behavior indicates that increasing the reflectivity does not necessarily improve the timing performance, even when it increases the number of detected photons. Multiple reflections increase the probability of photon detection but can also introduce a broader distribution of optical path lengths and arrival times. Therefore, the optical boundary conditions play a critical role in determining the ITR and must be considered together with photon statistics and detector geometry.\\
Overall, the results demonstrate that the ITR of the hexagonal BC404 scintillator is primarily influenced by the interaction configuration, while no significant dependence on the incident-muon energy was observed over the energy range investigated. In particular, the number of detected photons alone is not a sufficient metric for predicting timing performance, since the temporal distribution of the detected photons and their optical propagation paths also contribute significantly to the ITR.\\
These findings are relevant to the design and optimization of fast-timing detectors based on plastic scintillators and SiPM readout, particularly for applications such as cosmic-ray detection and time-of-flight systems. Future studies should incorporate a more realistic photosensor response, including photon detection efficiency, electronic noise, and timing jitter, as well as alternative detector geometries and reflective coatings. Such extensions would provide a basis for experimental validation and for further optimization of the timing performance of scintillator-based detectors.

\printbibliography
\end{document}